\documentclass{epl}

\usepackage{graphicx,epsfig}
\usepackage{dcolumn}
\usepackage{bm}
\usepackage{amssymb}

\title{Temporal Effects of Agent Aggregation 
in the Dynamics of a Competing Population}
\author{C. H. Yeung and K. Y. Michael Wong}
\institute{
Department of Physics, Hong Kong University of Science and Technology, 
Clear Water Bay, Hong Kong, China
}


\begin{document}

\maketitle

\begin{abstract}
We propose a model of a competing population 
whose agents have a tendency to balance their decisions in time. 
We find phase transitions to oscillatory behavior, 
explainable by the aggregation of agents into two groups. 
On a longer time scale, we find that 
the aggregation of smart agents 
is able to explain the lifetime distribution of epochs 
to 8 decades of probability.
\end{abstract}




Many natural and artificial systems involve interacting agents, 
each making independent decisions to pursue their own short-term objectives, 
but globally exhibiting long-time correlations 
beyond the memory sizes or the action cycles of the individuals 
\cite{anderson1988,schweitzer2002,challet2005}. 
Examples include the price fluctuations in financial markets 
\cite{stanley1995}, 
traffic patterns on the Internet \cite{traffic}, 
and the lifespan and metabolic time scales of living organisms. 
It is interesting to consider the extent to which 
the inherent properties of the systems can affect the temporal behavior.
Correlations beyond individual memory sizes 
can be consequences of interaction among agents. 
For example, some agents may find it advantageous 
to correlate their decisions over an extended period of time, 
leading to the emergence of epochs dominated by a group. 
This constitutes a system with multiple time scales. 
On a long time scale, it is described by the ages of the epochs, 
besides the shorter time scale of the underlying dynamics 
(usually determined by the memory sizes or the action cycles of the agents).

In this Letter, we analyze a competing population model 
with the above features. 
The model is most applicable to financial markets 
in which agents trade with finite capital, 
but can be extended to other multiagent systems 
such as routers in communication networks 
attempting to transmit multiclass traffic in a fair way. 
In an effort to avoid overdrawing her capital, 
an agent making a buying decision at an instant 
is more inclined to make a subsequent selling decision, and vice versa. 
This creates a natural short-time scale 
for the buy-and-sell cycles of the market. 
To model the adaptive attributes of the agents, 
each agent can occasionally revise her decision 
by following some adaptive strategy 
she perceives to be successful, 
analogous to the Minority Game (MG) \cite{challet2005}. 
We are interested in studying 
how agents aggregate in their decision making process 
in the short as well as long time scales.

Specifically, we consider a population of $N$ agents 
competing to maximize their individual utility, 
while striving to maintain a balance of their capital, 
$N$ being odd.
At each time step $t$, 
agent $i$ makes a bid $b_i(t)=1$ or $0$, 
and the minority group among the $N$ agents wins.
``1'' and ``0'' may represent ``buy'' and ``sell'' decisions 
in a financial market. 
To model agents with concerns for both finite capital and adatpive strategies, 
agent $i$ makes buying bids with probabilities $P[b_i(t)=1]$, where
\begin{equation}
	P[b_i(t)=1]=kf(c_i(t))+(1-k)P_{\rm strategy}[b_i(t)=1],
\label{bidprob}
\end{equation}
where $0\le k\le 1$ is the {\it restoring strength} 
determining the statistical weights of the two factors in Eq.~(\ref{bidprob}).

The first term models the consideration of finite capital 
in an agent's buying decisions at time $t$, 
whose probabilities increase with her capital $c_i(t)$ at the instant. 
For example, we choose in this Letter 
a three-step function $f(c_i(t))$ 
which is equal to 0, 1/2 and 1 respectively in the ranges 
$0\le c_i(t)<c_<$, $c_<\le c_i(t)<c_>$, and $c_>\le c_i(t)\le 1$.
For a buying bid, the capital reduces by $(1-\alpha)c_i(t)$, 
in exchange for the same value of stocks 
(to the lowest order approximation, 
we have neglected price changes due to fluctuations in supply and demand). 
Similarly, for a selling bid, 
the volume of stocks reduces by $(1-\alpha)[1-c_i(t)]$, 
in exchange for the same value of capital. 
Thus, the capital change is given by
\begin{equation}
	c_i(t+1)=\alpha c_i(t)+(1-\alpha)[1-b_i(t)].
\label{capital}
\end{equation}
In Eq.~(\ref{capital}), the second term 
counts the number of selling bids in history 
at a value of $1-\alpha$, 
and the first term shows that its effect on the capital 
is discounted exponentially at a rate of $\alpha$ per step. 
Hence, $\alpha$ is referred to as the {\it restoring memory factor}.
Small $\alpha$ corresponds to short-term restoring,
namely, balancing buying and selling in small number of steps,
whereas large $\alpha$ corresponds to long-term restoring.
When there are more selling (buying) bids in the recent bid history,
the capital tends to be high (low).
Through the probability $f(c_i(t))$,
the agent tends to balance the more frequent recent bids
with the opposite actions.
This is equivalent to adding a restoring force
to the decision making process of the agents.

The second term models an agent's tendency to use adaptive strategies,
and the probability $P_{\rm strategy}[b_i(t)=1]$
can be determined by standard games
such as the MG \cite{challet2005}
or the Evolutionary Minority Game (EMG) \cite{johnson1999a}.
In MG, strategies are binary functions
mapping an $m$-bit signal to an output bit, 
the signal here being the winning bits of the most recent $m$ steps. 
Before the game starts,
each agent randomly picks $s$ strategies. 
At each time step, the {\it cumulative payoffs} of strategies
that give a correct prediction increase by 1, 
while those predicting incorrectly decrease by 1.
Each agent then follows 
the strategy with the highest cumulative payoff among those she owns.
The EMG differs from the MG in details, but shares the same feature that the 
adopted strategies evolve according to historical performances.

The system behaviour is characterized
by the variance of buyers at the steady state. 
In the simulations, 
all agents initially hold equal value of capital and stocks, 
and the cumulative payoffs of strategies in MG
are randomized with variance $R$ \cite{wong2004}.
Remarkably, Fig.~1(a) shows that the system behaviour crosses over from a 
{\it strategy-specific} regime at $k=0$ to a {\it memory-factor-specific}
regime when $k$ approaches 1. In the latter regime,
the variances approach values dependent on $\alpha$,
{\it irrespective of the type of the adaptive strategies}
(such as in MG or EMG).
Thus in the strongly restoring limit,
the system behavior remains the same for a given value of $\alpha$,
as long as {\it some} adaptive strategy is adopted.
For large $\alpha$ in the strongly restoring limit,
the restoring force tends to bring the variance of the system
to the coin-toss limit $\sigma^2/N=1/4$,
implying that the agents make random and uncorrelated decisions.
For small $\alpha$ in the same limit,
the variance has a value lower than the coin-toss limit.
Note that the variances in the limit $k$ approaching 1
jumps discontinuously to the coin-toss limit at $k$=1
(that is, when the decisions are solely based
on the concern for finite capital),
showing that adaptive strategies,
even only rarely used by the agents,
are essential in bringing the system to an efficient state.

\begin{figure}
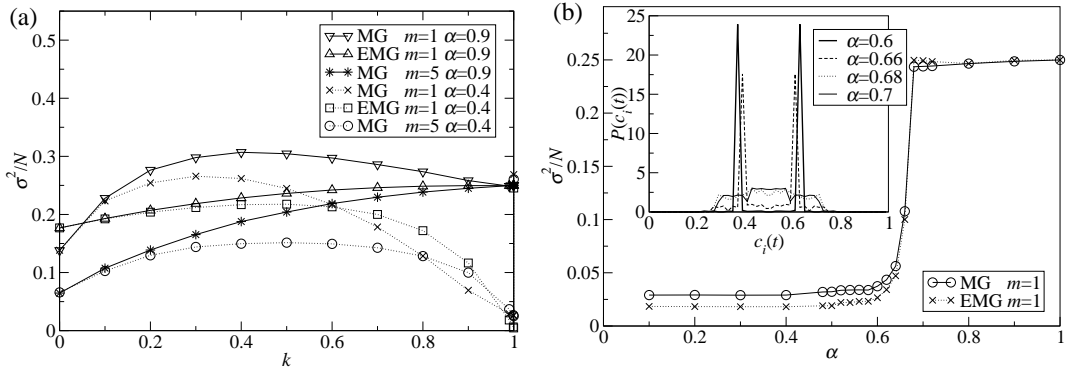

\leftline{\epsfig{figure=varAtt1VsK_bw.eps,width=0.48\linewidth}}
\vspace{-4.9cm}
\leftline{\hspace{7cm}
\epsfig{figure=varVsAlpha_CapDis_bw.eps,width=0.49\linewidth}}
\caption{(a) The variance as a function of the restoring strength $k$ 
($N=101$, 50,000 steps, 1,000 samples; for MG, $R=51$, $s=2$). 
(b) The variance as a function of the restoring memory factor $\alpha$ 
at $k=0.99$ and $\alpha_c=2/3$.
Inset: The distribution of capital for different $\alpha$ 
indicated in the legend.
}
\vspace{-0.4cm}
\end{figure}

Figure 1(b) shows the variance in the strongly restoring limit,
which increases with the restoring memory factor $\alpha$.
It undergoes a continuous phase transition to the coin-toss limit
at a critical $\alpha$,
whose value is independent of the type of adaptive strategies.
The nature of this transition is revealed 
in the distribution of capital
at a steady-state instant of the system, 
which has two sharp peaks in the short-term memory phase,
but one otherwise (Fig. 1(b) inset). 
The two-peaked distribution arises from the intentions of the agents
to balance their budget in a small number of steps,
causing them to aggregate into two groups, A and B,
analogous to the crowd-anticrowd picture of multiagent systems
\cite{johnson1999b}.
Group A consists of agents making alternating buy and sell bids
at odd and even time steps respectively,
and agents in group B make opposite alternating bids.
This self-organization of agent behavior is further confirmed
by the spectrum of a Fast Fourier Transform
of the capital of a typical agent,
which shows a sharp peak at the frequency 0.5.
In fact, one can show from Eq.~(\ref{bidprob})
that when $\alpha<\alpha_c\equiv \min[c_</(1-c_<),(1-c_>)/c_>]$,
a limit cycle of period 2 can be formed
from the alternating bids of an agent. 
This enables the capital of an agent to avoid
staying in the intermediate region $(c_<,c_>)$ of random bids, 
resulting in a {\it segregated} phase with two groups of agents.
The value of $\alpha_c$ coincides
with the phase transition point in Fig.~1(b).
In contrast, for $\alpha>\alpha_c$,
there appears a growing fraction of agents in the region of random bids,
and the period-2 dynamics is not sustained,
resulting in a {\it clustered} phase with a coin-tossed variance.
We note in passing that segregated and clustered behavior
have been observed in a model of EMG
due to a different cause \cite{hod2002},
but the issue of time scales has not been studied.

We now consider temporal effects beyond the period-2 oscillations, 
focusing on the limit $k$ approaching 1 in the segregated phase.
At each time step, an average of $(1-k)N\equiv\tilde k$ agents
make decisions according to adaptive strategies, 
and thus have a chance to switch from group A to B, or vice versa.
Then, groups A and B change from majority to minority, or vice versa,
in a time scale longer than the period-2 oscillations.
The emergence of these multiple time scales is illustrated in Fig.~2,
in which the buyer population is essentially oscillating with period-2.
Occasional phase slips in the buyer population
signify switches from group A being the winners 
to group B, or vice versa.
This corresponds to instants in Fig.~2
where the population of group A
crosses the minority-majority boundary.
The lifetime of an {\it epoch} is the duration
for which group A or B remains winning continuously.

\begin{figure}
\centerline{\epsfig{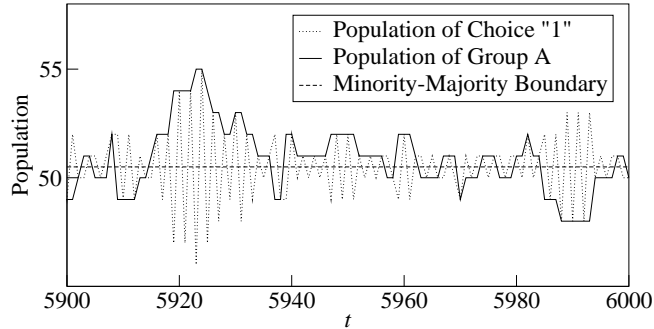}}
\vspace{-0.3cm}
\caption{
The population size of the buyers and group A 
for a particular sample 
($N=101$, $m=1$, $s=2$, $R=51$, $k=0.99$, $\alpha=0.4$, 10,000 steps). 
}
\vspace{-0.4cm}
\end{figure}

To analyze the variance and the epoch lifetimes, 
we note that since the history is dominated by period-2 oscillations, 
there is a reduction in the effective dimensionality of the dynamics. 
This greatly simplifies the analysis of the dynamics, 
rendering it generally applicable to MGs of arbitrary memory size $m$. 
Indeed, the two most common signals received by the agents 
are the series of alternating ``0'' and ``1'' 
(e.g. ``0101'' and ``1010'' in the case of $m=4$), 
referred to as the {\it epoch steady signals} (ES signals). 
After the first $m-1$ steps since their birth, 
the epochs enter the {\it epoch steady state} (ES state), 
in which the switching of agents between groups A and B 
are determined by their responses to the ES signals. 
Thus, there are 4 categories of strategies 
according to their response to the ES signals: 
{\it followers}, whose decision follows the last bit of the ES signals; 
{\it sellers}, whose decision is 0; 
{\it buyers}, whose decision is 1; 
{\it contrarian}, whose decision is opposite 
to the last bit of the ES signals.

For small $\tilde k$ such that $\langle L\rangle\gtrsim m$, the ES signals 
are the most frequent ones, 
causing the contrarian strategies 
to acquire the highest cumulative payoffs,
and the follower strategies the lowest.
The buyer and seller strategies predict correctly 
half of the times in the ES state, 
and their cumulative payoffs are intermediate.
At the ES state,
the agents making minority strategic decisions 
are those holding at least one contrarian strategy. 
They are referred to as {\it smart} agents. 
The probability of finding a smart agent is $1-(3/4)^s$.
Otherwise, if agents have at least one of the buyer or seller strategy, 
their strategic bid follows the minority decision in the ES state 
with probability 1/2, 
and are referred to as {\it mediocre} agents. 
The probability of this case is $(3/4)^s-(1/4)^s$.
With probability $(1/4)^s$,
the agents only have follower strategies at hand, 
and their strategic bid follows the majority decision in the ES state. 
They are referred to as the {\it dump} agents. 
On average, the probability of an agent 
bidding for the minority decision in the ES state 
is $T_{\rm ES}=1-(3/4)^s/2-(1/4)^s/2$.

The analysis of the first $m-1$ steps, 
referred to as the {\it epoch transient state} (ET state), 
is more involved. 
In this case, the smart agents 
are no longer receiving the ES signals 
that they are most adapted to. 
In particular, for those smart agents 
who hold only one contrarian strategy, 
the payoff of the contrarian strategy is accumulated 
owing to its successful responses to the dominant ES signals, 
irrespective of their responses to the ET signals. 
Thus, their average probability of making a minority decision is 1/2. 
On the other hand, for those smart agents 
who hold more than one contrarian strategies, 
the payoffs of the contrarian strategies 
are accumulated to the same extent in the ES state, 
and the ET signals are able to favour the payoff accumulation 
of those contrarian strategies 
that predict the correct minority bit in the ES state. 
These agents, who adapt to the ES state 
as well as partially to the ET state, 
are referred to as the {\it super-smart} agents. 
Their probability of making a minority decision is $T_{\rm super}$, 
whose average value can be shown to be greater then 1/2 \cite{yeung2006}, 
using a self-consistent and semi-empirical approach 
through an analysis of the historical occurrence of signals, 
or an approximation that all ET signals 
have the same probability of occurrence. 
Similarly, there are {\it super-mediocre} agents 
who hold more than one buyer or seller strategies. 
Since all dump agents hold more than one follower strategies, 
they can also be called {\it super-dump} agents. 
Their average probability of making a minority decision 
is also $T_{\rm super}$, 
in contrast to the probability of 1/2 
for the rest of the population. 
On average, the probability of an agent 
bidding for the minority decision in the ET state is 
$T_{\rm ET} = f_{\rm super}T_{\rm super} + (1-f_{\rm super})/2$, 
where $f_{\rm super}=1-(s/4)(3/4)^{s-1}-(s/2)(1/4)^{s-1}$.

We first analyze the dynamics 
in the {\it mean-rate} approximation.
The average probability
that an agent uses an adaptive strategy and switches side at time $t$ 
is, respectively, $(1-k)T(t)$ from majority to minority,
and $(1-k)(1-T(t))$ in the opposite direction,
where $T(t)$ = $T_{\rm ES}$ and $T_{\rm ET}$ 
in the ES and ET states respectively. 
The master equation for the distribution 
of the number $N_{\rm_A}$ of agents in group A
can then be solved numerically or by Monte Carlo.

As shown in Fig.~3, the average epoch lifetime $\langle L\rangle$
decreases with the average number $\tilde k$ of agents
using adaptive strategies.
When $\tilde k$ approaches 0, $\langle L\rangle$ appears to approach the 
expected result of $\langle L\rangle\sim\tilde k^{-1}$. This implies that 
during the average lifetime of an epoch, an average of $\tilde k\langle L
\rangle \approx 5$ agents make strategic decisions which may result in 
switching sides.
Furthermore, the prediction of the mean-rate approximation
has an excellent agreement with the simulation results.
A similarly excellent agreement is shown in Fig.~3 inset 
for the variance $\sigma^2$.
When $\tilde k$ approaches 0, the master equation for $P(N_{\rm_A})$
can be approximated by a Markov chain, yielding 
$P(N_{\rm A})\sim[(1-T_{\rm ES})/T_{\rm ES}]^{|N_{\rm_A}-(N+1)/2|}$
and a variance of 
$2[(1-T_{\rm ES})/(2T_{\rm ES}-1)]^2
+2[(1-T_{\rm ES})/(2T_{\rm ES}-1)]+1/4$.

\begin{figure}
\centerline{\epsfig{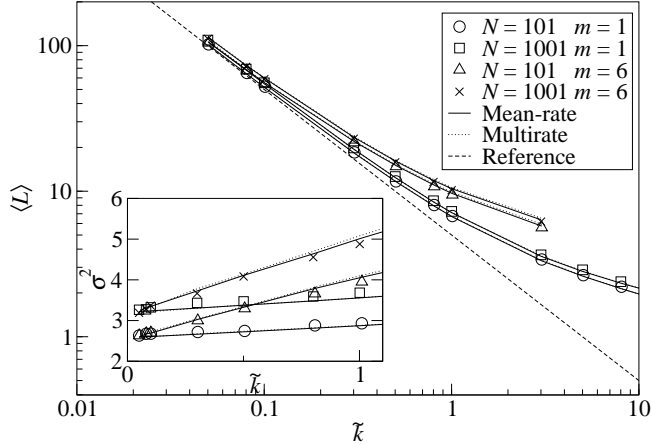}}
\vspace{-0.3cm}
\caption{Lifetime $\langle L\rangle$ as a function of $\tilde k$. 
($s=2$, $R=51$, $\alpha=0.4$, 30,000 steps with 2,000 samples) 
compared with the mean-rate and multirate approximations.
Reference: $\langle L\rangle=5\tilde k^{-1}$.
Inset: Variance $\sigma^2$ as a function of $\tilde k$.
}
\vspace{-0.4cm}
\end{figure}

Figure 4 shows that the epoch lifetimes 
follow an exponential distribution
near its maximum at small $L$.
The exponential decay rate depends on $\tilde k$
and is insensitive to $N$.
The mean-rate approximation has an excellent agreement
with simulation results in this regime.
When $L$ increases further,
the distribution crosses over to an exponential one 
with a lower decay rate dependent on both $\tilde k$ and $N$.
However, the mean-rate approximation predicts a {\it lower} distribution
of long epochs than simulations.
As shown in Fig.~4 inset, the decay rates of both the simulation and 
the mean-rate approximation approaches the same asymptotic value 
for large $N$, 
but their differences are significant in the mesoscopic regime. Analysis 
shows that the differences scale as $N^{-1.1\pm0.1}$.

The discrepancy arises from the inability of the mean-rate approximation
to differentiate the aggregation effects of various agents 
in the majority and minority groups.
Suppose it happens that the majority group in an epoch
consists of less smart agents than the minority group.
It follows that when an agent in the majority group
chooses to make a strategic decision,
it is less likely that she switches to the minority group.
The converse is valid for dump agents,
but to a lesser extent due to their smaller population.
This reduces the flow of agents from the majority to the minority group,
and hence lengthen the lifespan of an epoch.
This effect is most significant in the mesoscopic regime when the fluctuations
of the smart agent aggregate and the minority group size are important.

\begin{figure}
\centerline{\epsfig{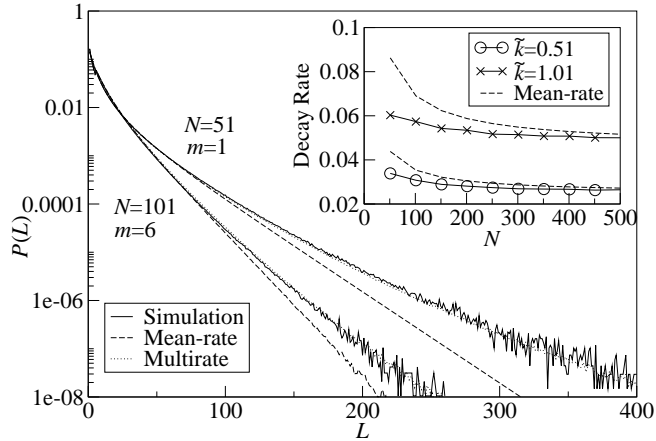}}
\vspace{-0.3cm}
\caption{Lifetime distribution ($s=2$, $R=51$, 
$k=0.99$, $\alpha=0.4$, 30,000 steps with 30,000 samples) 
compared with the mean-rate and multirate approximations.
Inset: decay rate of long epochs as a function of $N$ for $m=1$. 
}
\vspace{-0.4cm}
\end{figure}

The evidence that smart agents aggregate in the minority group
at the expense of the majority group is demonstrated in Fig.~5. 
We measure the difference in the number of smart agents
between the minority and majority groups ($N_{\rm min,s}-N_{\rm maj,s}$) 
averaged over an epoch.
Compared with the distribution for short epochs,
the peak of the distribution for long epochs
is shifted to the positive side.
This shows that long epochs have more smart agents
distributed in the minority group on average.

This leads to the second approximation of our analysis,
the {\it multirate} approximation.
We denote by $N_{\rm As}$, $N_{\rm Am}$ and $N_{\rm Ad}$
the number of smart, mediocre and dump agents in group A respectively.
Instead of using a single average probability 
of switching from the majority to the minority group 
for all agents and vice versa, 
the average transition probabilities for each of these agent types 
is taken into account. 
The master equation for the probability $P(N_{\rm As},N_{\rm Am},N_{\rm Ad})$
can be solved by Monte Carlo.
As shown in Fig.~4, the multirate approximation
yields a significantly higher probability for long epochs
compared with the mean-rate approximation,
resulting in an excellent agreement with simulation results
over 8 decades of probability.
Furthermore, the distribution of smart agents
derived from the multirate approximation
has an excellent agreement with simulation results,
as shown in Fig.~5.
This shows that the aggregation of smart agents
is crucial in explaining the occurrence of long epochs.

\begin{figure}
\centerline{\epsfig{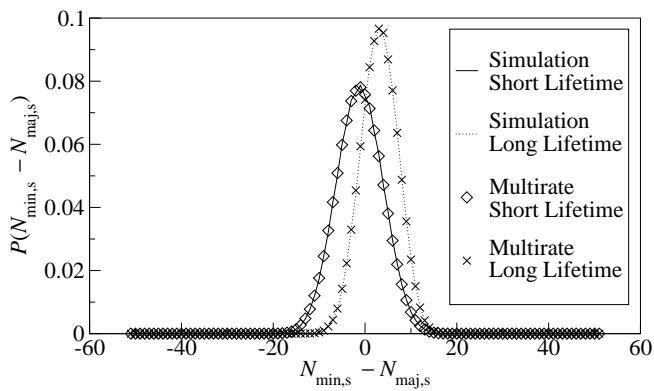}}
\vspace{-0.3cm}
\caption{Distribution of the difference ($N_{\rm min,s}-N_{\rm maj,s}$) 
between the number of smart agents in the minority and majority groups 
($N=51$, $m=6$, $s=2$, $R=51$, $\tilde k=0.51$, $\alpha=0.4$, 
30,000 steps with 30,000 samples).
The average lifetime $L\approx 11$. 
Epochs with short and long lifetimes are defined 
by $L\le 20$ and $L\ge 100$ respectively.
}
\vspace{-0.4cm}
\end{figure}

Due to their generic nature,
effects of smart agent aggregation
are expected to be present for adaptive strategies
other than MG strategies.
The essence is the stabilization of the minority group
by the smart agents,
whose strategies are favored by the sustenance of an epoch.
Note that the smart agents do not have any intrinsic preference
to either group A or B;
they favor the group they aggregate in
as long as it is the minority.
Differences among the various adaptive strategies
only come from details of the transition rate
between the majority and minority groups.

In summary, we have proposed a model of a competing population 
with multiple time scales.
When the tendency to balance the decisions of the agents is strong,
changing their intentions from long-term to short-term
induces phase transitions
to oscillations with two groups of agents
making alternating but opposite decisions,
and the behavior is independent of the details of the adaptive strategies.
On a longer time scale,
the history is characterized by epochs
dominated by one of the two groups.
Epochs end when sufficient numbers of agents
follow adaptive strategies and switch to the winning side.
Epochs are stabilized by the aggregation of smart agents
on the winning side in the mesoscopic regime.
The study of epoch lifetime is important to the agents who want to stay on 
the winning side as long as possible, or at least to have a knowledge of how 
long they would prevail. 
These generic features are relevant to multiagent systems
such as financial markets of agents with finite capital, distributed 
control of multiclass traffic in communication networks, and other systems 
with competing aggregates of agents.

We thank S. W. Lim, D. Saad and Y. P. Ma for discussions.
This work is supported by the Research Grant Council of Hong Kong 
(HKUST6062/02P and DAG04/05.SC25).


\end{document}